\documentclass[dvipdfmx]{article}

\usepackage{amssymb,amsfonts,amsmath,amsthm,stmaryrd}
\usepackage{cite,enumerate,float,indentfirst}
\usepackage[dvipdfmx]{graphicx}
\usepackage{tikz}
\usepackage{color}
\usepackage{bm}
\usepackage{simplewick}
\usepackage{here}

\theoremstyle{definition}

\theoremstyle{remark}

\allowdisplaybreaks

\def\be{\begin{eqnarray}}
\def\ee{\end{eqnarray}}

\def\p{\partial}

\def\Tr{{\rm Tr}\,}

\definecolor{red}{rgb}{1,0,0}
\definecolor{orange}{rgb}{1,0.5,0}
\definecolor{violet}{rgb}{0.7,0,1}

\def\cre{\color{red}}

\def\cg{\color{green}}
\def\cb{\color{blue}}

\hoffset= - 1.0in         

\begin{document}

\begin{center}
\begin{small}
\hfill NITEP 117  \\
\hfill OCU-PHYS 546 \\
\hfill FIAN/TD-11/21\\
\hfill IITP/TH-18/21\\
\hfill ITEP/TH-27/21\\
\hfill MIPT/TH-15/21\\
\end{small}
\end{center}

\hfill September, 2021

\bigskip

\begin{center}{\Large{\bf Review on the Operator/Feynman diagram/Dessins d'enfants \\
\vspace{.3cm}
Correspondence in Tensor Model}}
\end{center}

\bigskip

\centerline{\large H. Itoyama$^{1,2,3}$, A. Mironov$^{4,5,6}$, A. Morozov$^{7,5,6}$, and R. Yoshioka$^{3}$}

\vspace{1cm}

\noindent
{$^1$ \it Nambu Yoichiro Institute of Theoretical and Experimental Physics (NITEP)}\\
{$^2$ \it Department of Mathematics and Physics, Graduate School of Science, Osaka City University}\\
{$^3$ \it Osaka City University Advanced Mathematical Institute (OCAMI), \\
\phantom{aaaaaaaa} 3-3-138, Sugimoto, Sumiyoshi-ku, Osaka, 558-8585, Japan}\\
{$^4$ \it I.E.Tamm Theory Department, Lebedev Physics Institute, Leninsky prospect, 53, Moscow 119991, Russia}\\
{$^5$ \it A.I. Alikhanov  Institute for Theoretical and Experimental Physics of NRC “Kurchatov Institute”, \\
\phantom{aaaaaaaa} B. Cheremushkinskaya, 25, Moscow, 117259, Russia}\\
{$^6$ \it Institute for Information Transmission Problems of RAS (Kharkevich Institute),\\
\phantom{aaaaaaaa} Bolshoy Karetny per. 19, build.1, Moscow 127051 Russia}\\
{$^7$ \it Moscow Institute of Physics and Technology, Dolgoprudny, 141701, Russia}

\vspace{1cm}

\centerline{ABSTRACT}

\bigskip

{\footnotesize
}
A short review of the Operator/Feynman diagram/\textit{dessin d'enfants} correspondence in the rank 3 tensor model is presented,
and the cut \& join operation is given in the language of \textit{dessin d'enfants} as a straightforward development.
We classify operators of the  rank 3 tensor model up to level 5 with \textit{dessin d'enfants}\footnote{Based on the talk given by R. Y. at the international workshop
``Randomness, Integrability and Representation Theory in Quantum Field Theory" at Osaka City University Media Center on March 25, 2021.}.

\bigskip

\bigskip

\section{Introduction}
The tensor model \cite{David1985,ADJ1991,Sasakura1991,Gross1992}
(\cite{IMM1703,IMM1704,MM1706,IMM1710,IMM1808,IYoshi1903,IMM1909,IMM1910,MM2020} for our series of work, \cite{KPT1808} for a review) is a generalization of the rectangular complex matrix model:
 it is obtained by replacing the rectangular matrix with a tensor of rank $r$.
It has ``gauge" symmetry $U(N_1) \otimes \cdots \otimes  U(N_r)$.
The gauge invariant operators in the tensor model are constructed by all possible contractions of $n$ tensors $M_{a_1\cdots a_r}$ and $n$ conjugate tensors $\bar{M}^{a_1\cdots a_r}$,
where the number $n$ is called level.
In the  matrix model ($r=2$ case), any gauge invariant operator is realized as a product of trace operators, and therefore can be labeled by a partition,
 \begin{equation}
 \Tr(A\bar{A})^{n_1}\Tr(A\bar{A})^{n_2} \cdots~~~
 \leftrightarrow~~~(n_1,n_2, \cdots).
 \end{equation}
 Here $A$ and $\bar{A}$ are respectively the rectangular matrix and its conjugate.
The rank $r \geq 3$ tensor model, in contrast,  contains many non-trivial operators,
  and therefore the enumeration problem of the operators is more nontrivial  \cite{BGR1307,IMM1710,BG2005,BGR2106,MN2010}.
In the previous paper \cite{AIMMVY1911}, we have demonstrated the OP/FD/\textit{dessin} correspondence.
This correspondence suggests a classification of operators in  the rank $r$ tensor models by Feynman diagrams in the  rank $r-1$ tensor models and, in the  particular case of  rank $r=3$, by bipartite graphs called Grothendieck's \textit{dessins} (\textit{d'enfant}) \cite{LZ2004,JRRG1012,AADKK0710}.

On the other hand, there exist cut and join operations, which form basic structures of  the Virasoro-like constraints
 \cite{David1990,MM1990,AM1990,IM1991N,IM1991W}
 associated with the change of integration variables in the partition function \cite{IMM1710,IYoshi1903}.
They can also be viewed as operations that generate operators of the model
and, in principle, they can generate all operators starting from a properly chosen set of keystone operators.
Since the role of the join operation is to raise the level of the operator, the join operation generates a set of higher level operators from a given keystone operator, which is called a join pyramid.
There exists, however, an infinite number of operators which are not included in the join pyramid,
 and these can only be generated by the cut operation.
Since the cut operation is an operation that lowers the level of operator by one,
 the only way to find all missing operators appears to examine the operators one level higher, one by one.
Each of these operators newly obtained by the cut operation will form a new join pyramid by applying the join operation and this series of operations continues indefinitely.
The structure of the set of operators of this kind represents a part of the complexity of the operator classification problem.
See \cite{IYoshi1903} for a direction to try to relax this.

The purpose of this paper is to review the OP/FD/\textit{dessin} correspondence and
 to attempt to rewrite the join and cut operations in the language of \textit{dessin} for $r=3$.
In the next section,  we begin with reviewing the OP/FD/\textit{dessin} correspondence.
As a straightforward development of this correspondence, we give the classification of operators of the rank 3 tensor model in terms of the Feynman diagrams of the rectangular complex matrix model and of the \textit{dessins}, which we give explicitly up to level 5 in Tables 1-5 and in Appendix A.
In section three, we  investigate the behavior of the cut and join operations by \textit{dessins}.

Note that, in this review, we use the notation of operators following \cite{IMM1710}.

\section{\textit{Dessins d'enfants} and OP/FD/\textit{dessin} correspondence}
We start with reviewing the \textit{dessins d'enfant} \cite{LZ2004} and the OP/FD/\textit{dessin} correspondence.
The \textit{dessin} is a bipartite graph $D$ embedded into an orientable two-dimensional surface $X$.
The bipartite graph is a graph of two colored vertices and edges,
 where the edges connect only vertices of different colors.
 In this paper,  we choose red and green as color of the vertices.
 In addition, the complement  $X \backslash D$  has to be a disjoint union of the connected components called faces. Each face is homeomorphic to an open disk.

The \textit{dessin} can also be expressed by a set of  three permutations $[\sigma_r, \sigma_g,\sigma_b]$
 such that $\sigma_b \sigma_g \sigma_r = 1$,
which is called a 3-constellation.
Here $\sigma_r, \sigma_g,\sigma_b \in S_d$ for \textit{dessin} with $d$ edges.
The 3-constellation  can  be constructed as follows:
first, label all edges with distinct numbers from 1 to $d$.
Then, read the labels at the vicinity of each red vertex counter-clockwise around it,
this produces a cycle in the permutation $\sigma_r$, the whole $\sigma_r$ being a
set of such cycles associated with all red vertices. 
Since each edge is associated with only one red vertex, every edge number enters once, and this procedure is unambiguous.
Similarly, one can construct $\sigma_g$.
Finally, $\sigma_ b = (\sigma_g \sigma_r)^{-1}$.
See Appendix \ref{list}.

In \cite{AIMMVY1911}, we proposed a graphical method to understand the OP/FD/\textit{dessin} correspondence.

\bigskip

\noindent
\textbf{\underline{OP/FD}}

\vskip.5\baselineskip

Let us take the rank 3 operator $K_1$ as a simplest example, which is given by
\begin{equation}\label{K1:def}
  K_1 = M_{{\cre a_1}{\cg a_2}{\cb a_3}}\bar{M}^{{\cre a_1}{\cg a_2}{\cb a_3}} = ~
  \lower3ex\hbox{\begin{tikzpicture}
   \draw [thick,red] (1,0) arc (0:180:0.5) ;
   \draw [thick,green] (1,0) arc (0:-180:0.5) ;
   \draw[thick,blue](0,0)--(1,0);
   \filldraw [thick,fill=white] (0,0) circle (0.1);
   \filldraw [thick,fill=black] (1,0) circle (0.1);
   \end{tikzpicture}} ,
\end{equation}
where  the white and black dots are  the rank 3 tensor $M$ and its conjugate $\bar{M}$, respectively, and the colored lines represent the indices to be contracted.
In order to obtain the corresponding Feynman diagram,
we can reinterpret one of the indices, say blue, as a symbol representing a pair of tensors which are Wick contracted.
In the case of $K_1$,
\begin{equation}
  K_1 ~~ \longleftrightarrow~~
  \lower3ex\hbox{\begin{tikzpicture}
   \draw [thick,red] (1,0) arc (0:180:0.5) ;
   \draw [thick,green] (1,0) arc (0:-180:0.5) ;
   \draw[<->, thick,blue](0.1,0)--(0.9,0);
   \filldraw [thick,fill=white] (0,0) circle (0.1);
   \filldraw [thick,fill=black] (1,0) circle (0.1);
   \end{tikzpicture}} =
    \contraction{}{A}{{}_{{\cre a_1} {\cg a_2}}}{\bar{A}}
    A_{{\cre a_1} {\cg a_2}} \bar{A}^{{\cre a_1}{\cg  a_2}} = {\cre N_1} {\cg N_2},
\end{equation}
Here the dots are read as the rank 2 tensors (rectangular matrices) $A$ and $\bar{A}$ which are contracted by two colored indices.

The generalization to other operators is straightforward.
Consider any graph constructed by three colored (red, green, blue) edges,  then there exist two kinds of interpretation of the blue lines.
Blue has been chosen just to specify one of the three colors.
The first interpretation is that the blue lines are one of contractions of the indices.
This interpretation yields the rank 3 operator.
The second interpretation is that the blue lines determine the Wick contraction pair.
This interpretation yields a Feynman diagram of the corresponding rank 2 operator.
These are in correspondence with each other.

The OP/FD correspondence can be extended to the higher rank models as well:
 each operator in the rank $r$ tensor model has a corresponding Feynman diagram in the rank $r-1$ model.
Clearly their numbers match,
\begin{equation}
\mathcal{N}_r(n) = \textbf{N}_{r-1}(n).
\end{equation}

\noindent
\textbf{\underline{OP/\textit{dessin}}}

\vskip.5\baselineskip

Let us take $K_1$ again.
As can be seen in eq.\eqref{K1:def}, $K_1$ has  a red-blue cycle and a green-blue cycle.
After we paint the areas surrounded by these cycles red and green, respectively, we obtain the two surfaces colored respectively by red and green, called red and green faces, with the blue line as the common boundary,
\begin{equation}
K_1 = \lower3ex\hbox{\begin{tikzpicture}.
   \draw [thick,red] (1,0) arc (0:180:0.5) ;
   \draw [thick,green] (1,0) arc (0:-180:0.5) ;
   \draw[thick,blue](0,0)--(1,0);
   \filldraw [thick,fill=white] (0,0) circle (0.1);
   \filldraw [thick,fill=black] (1,0) circle (0.1);
   \end{tikzpicture}}
   ~~ \longleftrightarrow~~
\lower3ex\hbox{\begin{tikzpicture}
 \filldraw [draw=red, fill=red] (1,0) arc (0:180:0.5) ;
 \filldraw [draw=green, fill=green] (1,0) arc (0:-180:0.5) ;
 \draw[ultra thick,blue](0,0)--(1,0);
 \end{tikzpicture}}~ .
\end{equation}
Taking the dual of this surface,
\begin{equation}\label{op-dessin}
\begin{tabular}{ccc}
\text{operator} && \text{\textit{dessin}} \\
red face&$\leftrightarrow$&red vertex \\
green face&$\leftrightarrow$&green vertex \\
blue boundary&$\leftrightarrow$&edge connecting bicolored vertices
\end{tabular}
\end{equation}
we obtain the corresponding \textit{dessin}:
\begin{equation}
K_1
   ~~ \longleftrightarrow~~
\lower3ex\hbox{\begin{tikzpicture}
 \filldraw [draw=red, fill=red] (1,0) arc (0:180:0.5) ;
 \filldraw [draw=green, fill=green] (1,0) arc (0:-180:0.5) ;
 \draw[ultra thick,blue](0,0)--(1,0);
 \end{tikzpicture}}
    ~~ \longleftrightarrow~~
  \lower3ex\hbox{\begin{tikzpicture}
    \draw[thick] (0,0.4) -- (0,-0.4);
    \filldraw [thick,fill=red] (0,0.4) circle (0.1);
    \filldraw [thick,fill=green] (0,-0.4) circle (0.1);
    \end{tikzpicture}}~ .
\end{equation}
Similarly, any operator of the rank 3 tensor model has a corresponding \textit{dessin}.
The number of edges in a \textit{dessin} is equal to the level of the corresponding operator because the number of  blue lines is the same as the level of the operator, and
the blue lines are in one-to-one correspondence with the edges of the \textit{dessins}.
It is known \cite{IMM1710} that the numbers of connected operators up to level 5 are 1, 3, 7, 26, 97, respectively, which are exactly equal to the numbers of \textit{dessins} with a given number of edges.

\begin{figure}[H]
  \centering
  \includegraphics[height=5.5cm]{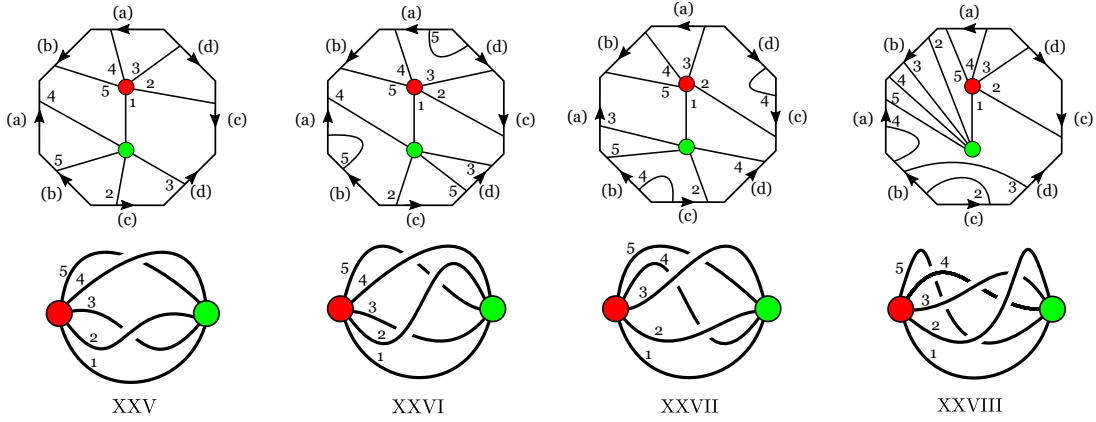}
  \caption{XXV = [(12345),(14523),(12453)], XXVI = [(12345),(14253),(12345)], XXVII = [(12345),(13524),(13524)], XXVIII=[(12345),(12345),(14253)]
  }
  \label{fig:genus2}
 \end{figure}

As another example, let us take a look at  $K_{3W}$, which is given by
\begin{equation}
 K_{3W} = ~ \lower5ex\hbox{\begin{tikzpicture}
     \draw[thick,blue](0,0)--(1.6,0);
     \draw[thick,blue](0.4,0.7)--(1.2,-0.7);
     \draw[thick,blue](0.4,-0.7)--(1.2,0.7);
     \draw[thick,red](0,0)--(0.4,0.7);
     \draw[thick,green](0.4,0.7)--(1.2,0.7);
     \draw[thick,red](1.2,0.7)--(1.6,0);
     \draw[thick,green](1.6,0)--(1.2,-0.7);
     \draw[thick,red](1.2,-0.7)--(0.4,-0.7);
     \draw[thick,green](0.4,-0.7)--(0,0);
      \filldraw [thick,fill=black] (0,0) circle (0.1);
      \filldraw [thick,fill=black] (1.2,0.7) circle (0.1);
      \filldraw [thick,fill=black] (1.2,-0.7) circle (0.1);
      \filldraw [thick,fill=white] (1.6,0) circle (0.1);
      \filldraw [thick,fill=white] (0.4,0.7) circle (0.1);
      \filldraw [thick,fill=white] (0.4,-0.7) circle (0.1);
 \end{tikzpicture}} ~.
\end{equation}
By rearrangement of the white and black dots, this operator can be transformed as follows:
\begin{equation}
\includegraphics[height=2.2cm]{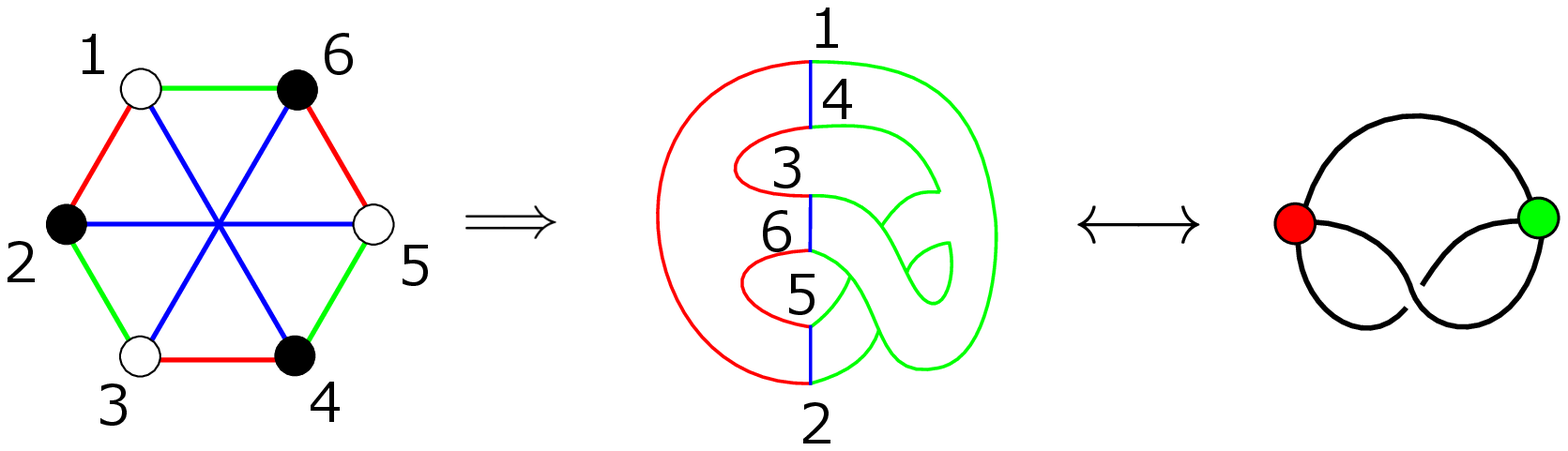}
\end{equation}
The resulting \textit{dessin} is clearly nonplanar and can not be drawn on sphere without self-intersections but only on torus.
Here the Riemann-Hurwitz formula can be exploited:
\begin{equation}\label{RH}
2g-2 = d - { V_r} - { V_g} - { F},
\end{equation}
where $g$ is the genus of the surface on which the \textit{dessin} is drawn,
$d$ is the number of edges, $V_r$ and $V_g$ are respectively the numbers  of red and green vertices,
and $F$ is the number of  faces.
Note that $d$ is equal to the level of the corresponding operator.
Since $V_r, V_g, F \geq 1 $, the maximum genus $g_m$ at level $d$ is given by
 \begin{equation}
  g_{m} = \left\lfloor \frac{d-1}{2} \right\rfloor.
 \end{equation}
 Hence, a \textit{dessin} on genus $g$ surface can only appear at level $2g+1$ and above.
As mentioned above, the number of operators at level 3 is 7.
One of the 7 operators is $K_{3W}$, which is the only one that can be put on torus at level 3.
We have checked all operators up to level 5.
At level 4, 6 operators are on torus. At level 5, 33 operators are on torus, and 4 operators are on genus two surface, the latter being listed in Fig. \ref{fig:genus2}.

\noindent
\textbf{\underline{FD/\textit{dessin}}}

\vskip.5\baselineskip

The \textit{dessins} and the Feynman diagrams for rank two tensors are dual to each other, and
there is the following correspondence:
\begin{align*}
\begin{tabular}{ccc}
\textit{\textit{dessin}} && \text{Feynman diagram} \\
 face &$\leftrightarrow$ & rank 2 operator\\
edge&$\leftrightarrow$&propagator \\
red vertex&$\leftrightarrow$&red loop \\
green vertex&$\leftrightarrow$&green loop
\end{tabular}
\end{align*}
For example,
\begin{equation}
 (K_1 ~ \leftrightarrow )  ~~
 \lower3ex\hbox{\begin{tikzpicture}
    \draw[thick] (0,0.4) -- (0,-0.4);
    \filldraw [thick,fill=red] (0,0.4) circle (0.1);
    \filldraw [thick,fill=green] (0,-0.4) circle (0.1);
    \end{tikzpicture}} ~=~
\lower4ex\hbox{\begin{tikzpicture}
    \draw[thick] (0,0.4) -- (0,-0.4);
    \filldraw [thick,fill=red] (0,0.4) circle (0.1);
    \filldraw [thick,fill=green] (0,-0.4) circle (0.1);
    \draw[red, thick] (-0.3,0.1) rectangle (0.3,0.7);
    \draw [-latex,red] (0.2,0.1) -- +(0:0.1);
    \draw[green, thick] (-0.3,-0.1) rectangle (0.3,-0.7);
    \draw [-latex,green] (0.2,-0.1) -- +(180:0.1);
    \node at (0.5,0) {$\otimes$} ;
    \end{tikzpicture}}
  ~ \leftrightarrow ~
\lower3ex\hbox{\begin{tikzpicture}
    \draw [red, thick] (0,0) circle (0.3);
    \draw [green, thick] (0,0) circle (0.5);
    \draw [-latex,red] (0,-0.3) -- +(0:0.1);
    \draw [-latex,green] (0,-0.5) -- +(180:0.1);
    \node at (0.4,0) {$\otimes$} ;
    \end{tikzpicture}}
    =  \contraction{}{A}{{}_{{\cre a_1} {\cg a_2}}}{\bar{A}}
         A_{{\cre a_1} {\cg a_2}} \bar{A}^{{\cre a_1}{\cg  a_2}},
\end{equation}
where the rank 2 operator $ A\bar{A}$ is denoted by $\otimes$.

\vspace{10mm}

We have obtained all \textit{dessins} up to level 5, which are listed in Tables \ref{level1}-\ref{level5}.
For the notation of each operator, we refer the readers to \cite{IMM1710}. See also Appendix \ref{list}.
The superscripts in Table 5 and in Appendix A label operators obtained by permutations of colors.
The columns are classified by the numbers of two colored vertices $(V_r, V_g)$ and
the rows are classified by the partition representing the rank 2 operator that yields  the  corresponding Feynman diagram.
According to the FD/\textit{dessin} correspondence,
 the length of partition is equal to the number of  faces of \textit{dessin} $F$.
At odd levels, the \textit{dessins} with maximum genus appear in the first row and the first column in each table
because these are realized only when $V_r= V_g = F =1$.
\begin{table}[H]
\centering
\caption{level 1} \label{level1}
$\begin{array}{c|c}
    & (1,1) \\ \hline
  (1)& K_1
  \end{array}$
\end{table}
\begin{table}[H]
\centering
\caption{level 2} \label{level2}
$\begin{array}{c|cccc}
      & (1,1) & (1,2) & (2,1) \\ \hline
      (2)& & K_{\cre 2} & K_{\cg 2}  \\
      (1^2) & K_{\cb 2}  & &
    \end{array}$
\end{table}
\begin{table}[H]
\centering
\caption{level 3} \label{level3}
$\begin{array}{c|ccccccccc}
    & (1,1) & (1,2) & (2,1) &  (1,3) & (3,1) & (2,2)\\ \hline
    (3) & K_{3W} &   &  & K_{\cre 3}&K_{\cg 3}& K_{{\cre 2}{\cg 2}}\\
    (21)&&K_{{\cb 2}{\cre 2}}&K_{{\cg2 }{\cb 2}}&&&\\
    (1^3)&K_{\cb 3}&&&&&&
  \end{array}$
\end{table}
\begin{table}[H]
\centering
\caption{level 4} \label{level4}
${\footnotesize
  \begin{array}{c|cccccccc}
    & (1,1)&(1,2)&(2,1)&(1,3)&(3,1)&(1,4)&(4,1)\\ \hline
    (4)&&K_{\cre 22W}, K_{\cre 31W}&K_{\cg 22W}, K_{\cg 31W}&&&K_{\cre 4}&K_{\cg 4}&\\
    (31) &K_{\cb 31W}&&&K_{{\cre 3}{\cb 2}}&K_{{\cg 3}{\cb 2}}&&\\
    (2^2) &K_{\cb 22W}&&&K_{{\cre 2}{\cb 2}{\cre 2}}&K_{{\cg 2}{\cb 2}{\cg 2}}&&\\
    (21^2)&&K_{{\cb 3}{\cre 2}}, K_{{\cb 2}{\cre 2}{\cb 2}}&K_{{\cb 3}{\cg 2}}, K_{{\cb 2}{\cg 2}{\cb 2}}&&&&&\\
    (1^4)&K_{\cb 4}&&&&&&&\\
    &&&&&&&&\\
    &(2,2)&(2,3)&(3,2)&&&&&\\\hline
    (4)&&K_{{\cre 3}{\cg 2}}, K_{{\cre 2}{\cg 2}{\cre 2}}&K_{{\cg 3}{\cre 2}}, K_{{\cg 2}{\cre 2}{\cg 2}}&&&&&\\
(31)&K_{222}, K_{{\cg 2}{\cre 2}{\cb 2}}, K_{{\cb 2}{\cg 2}{\cre 2}}&&&&&&& \\
(2^2)&K_{{\cre 2}{\cb 2}{\cg 2}}, K_{4C}&&&&&&
  \end{array} }$
\end{table}
\begin{table}[H]
\centering
\caption{level 5} \label{level5}
{\footnotesize
\centerline{ $ \begin{array}{c|ccccccccc}
& (1,1) &(1,2)&(2,1)&(1,3)&(3,1)&(1,4)&(4,1)&(1,5)&(5,1) \\\hline
(5) & \mathrm{XXV}, \mathrm{XXVI},&&&\mathrm{VI^{(1)}}, \mathrm{VIII^{(1)}},&\mathrm{VI}, \mathrm{VIII},&&&\mathrm{I^{(1)}}&\mathrm{I}\\
& \mathrm{XXVI}^{(1)},\mathrm{XXVI}^{(2)} &&&\mathrm{V}^{(1)}&\mathrm{V}&&&& \\
(41)&&\mathrm{XII}^{(4)},\mathrm{XIII}^{(4)},\mathrm{XVIII}^{(4)},&\mathrm{XII}^{(2)},\mathrm{XIII}^{(2)},\mathrm{XVIII}^{(2)},&&&\mathrm{II}^{(4)}&\mathrm{II}^{(2)}&& \\
&&\mathrm{XVIII}^{(5)},\mathrm{XIX}^{(2)}&\mathrm{XVIII}^{(3)},\mathrm{XIX}^{(1)}&&&&&&\\
(32)&&\mathrm{XII}^{(5)},\mathrm{XIII}^{(5)},\mathrm{XIV}^{(2)}&\mathrm{XII}^{(3)},\mathrm{XIII}^{(3)},\mathrm{XIV}^{(1)}&&&\mathrm{III}^{(4)}&\mathrm{III}^{(2)}&& \\
(31^2)&\mathrm{VIII}^{(2)}, \mathrm{X}^{(2)}&&&\mathrm{IV}^{(4)},\mathrm{VII}^{(2)}&\mathrm{IV}^{(2)},\mathrm{VII}^{(1)}&&&&\\
(2^21)&\mathrm{VI}^{(2)}&&&\mathrm{IV}^{(5)},\mathrm{V}^{(2)}&\mathrm{IV}^{(3)},\mathrm{V}^{(1)}&&&&\\
(21^3)&&\mathrm{II}^{(5)},\mathrm{III}^{(5)}&\mathrm{II}^{(3)},\mathrm{III}^{(3)}&&&&&&\\
(1^5)&\mathrm{I}^{(2)}&&&&&&&&\\&&&&&&&&&\\\hline
&(2,2)&(2,3)&(3,2)&(2,4)&(4,2)&&&&\\\hline
(5)&\mathrm{XII},\mathrm{XII}^{(1)},\mathrm{XIII}&&&\mathrm{II}^{(1)},\mathrm{III}^{(1)}&\mathrm{II},\mathrm{III}&&&&\\
&\mathrm{XIII}^{(1)},\mathrm{XIV},\mathrm{XVIII},&&&&&&&&\\
&\mathrm{XVIII}^{(1)},\mathrm{XIX}&&&&&&&&\\
(41)&&\mathrm{XXIX}^{(1)},\mathrm{XXX}^{(1)},&\mathrm{XXIX},\mathrm{XXX},&&&&&&\\
&&\mathrm{XXXI}^{(1)},\mathrm{XXXII}^{(1)},&\mathrm{XXXI},\mathrm{XXXII},&&&&&&\\
&&\mathrm{XXXIII}^{(1)},\mathrm{XXIX'}^{(1)}&\mathrm{XXXIII},\mathrm{XXIX'}&&&&&&\\
&&&&&&&&&\\
(32)&&\mathrm{XXX}^{(4)},\mathrm{XXXI}^{(4)}, &\mathrm{XXX}^{(2)},\mathrm{XXXI}^{(2)},&&&&&&\\
&&\mathrm{XXXIII}^{(4)},\mathrm{XXXIV}^{(1)},&\mathrm{XXXIII}^{(2)},\mathrm{XXXIV}&&&&&&\\
&&\mathrm{XXX'}^{(1)}&\mathrm{XXX'}&&&&&&\\
(31^2)&\mathrm{XXXI}^{(3)},\mathrm{XXXI}^{(5)},&&&&&&&\\
&\mathrm{XXXII}^{(2)},\mathrm{XXXIII}^{(3)}&&&&&&&&\\
&\mathrm{XXXIII}^{(5)},\mathrm{XXX'}^{(2)}&&&&&&&&\\
&&&&&&&&&\\
(2^21)&\mathrm{XXIX}^{(2)},\mathrm{XXX}^{(3)}&&&&&&&&\\
&\mathrm{XXX}^{(5)},\mathrm{XXXIV}^{(2)}&&&&&&&&\\
&\mathrm{XXIX'}^{(2)}&&&&&&&&
\\&&&&&&&&&\\\hline
&(3,3)&&&&&&&&\\\hline
(5)&\mathrm{IV},\mathrm{IV}^{(1)},\mathrm{V},\mathrm{VII}&&&&&&&&\\&&&&&&&&
  \end{array} $
  }}
\end{table}

\noindent
\textbf{\underline{Gaussian average}}

\vskip.5\baselineskip

The correspondence between the rank $r$ operators and the rank $r-1$ Feynman diagrams can be demonstrated at the level of the Gaussian averages.
Let us define a map $FD_i: \mathrm{OP}^{(r)} \to \mathrm{FD}^{(r-1)}$, where
$\mathrm{OP}^{(r)}$ is the set of rank $r$ operators, and $\mathrm{FD}^{(r-1)}$ is the set of rank $r-1$ Feynman diagrams.
The subscript $i$ refers to the index that specifies the Wick contraction pair.
We have chosen $i= $blue in the above $r=3$ case.
The Gaussian average of any operator $K$ behaves like
\begin{equation}\label{<K>}
\langle K \rangle = N_i^n \prod_{j \neq i}  N_j^{\ell_j} + \mathcal{O}(N_i^{n-1}),
\end{equation}
where $n$ is the level of $K$, and $\ell_j$ is a set of integers less than or equal to $n$.
On the other hand, the value of $FD_i(K)$ is
\begin{equation}\label{vFD(K)}
{\rm value}(FD_i (K)) = \prod_{j \neq i} N_j^{\ell_j}.
\end{equation}
In the case of $r=3$, this becomes
\begin{equation}\label{vFD(K)r=3}
{\rm value}(FD_{\rm blue} (K)) = N_{r}^{V_r} N_{g}^{V_g}.
\end{equation}
Note that, in eq.  \eqref{vFD(K)r=3}, the exponents are the numbers of two colored vertices of the \textit{dessin}.
Clearly, the following general relation is satisfied in eqs. \eqref{<K>} and \eqref{vFD(K)}:
\begin{equation}
{\rm value}(FD_i (K)) = \lim_{N_i \to \infty} \frac{1}{N_i^n} \langle K \rangle.
\end{equation}
For example,
 \begin{align}
 &\left\langle( \Tr A\bar{A})^2 \right\rangle
  = FD_{\rm blue} (K_{\cb 2}) + FD_{\rm blue}(K_1^2)
  = \lim_{{\cb N_b} \to \infty} \frac{1}{{\cb N_b}^2}
  \langle K_{\cb 2} + K_1^2 \rangle, \\
  &\left\langle \Tr (A\bar A)^2\right\rangle
  = FD_{\rm blue} (K_{\cre 2}) + FD_{\rm blue} (K_{\cg 2})
  = \lim_{{\cb N_b} \to \infty} \frac{1}{{\cb N_b}^2}
  \langle K_{\cre 2} + K_{\cg 2} \rangle.
 \end{align}

\section{Cut and join}
 In this section,  we reexpress  the cut \& join operation of the rank three tensor model in the language of
  \textit{dessins d'enfants}, making exploit the OP/\textit{dessin} correspondence.
Let us recall the definition of the join operation \cite{IMM1710}:
\begin{equation}
 \{ K_1, K_2 \} = \sum_{a_1, a_2, a_3} \frac{\p K_1}{\p M_{a_1a_2a_3}} \frac{\p K_2}{\p \bar{M}^{a_1a_2a_3}}.
\end{equation}
Obviously, $M_{a_1a_2a_3}$ and $\bar{M}^{a_1a_2a_3}$ are cut out accordingly from $K_1$ and from $K_2$, and their  respective indices are connected.
Let us focus on $K_1$.
 In the interior of $K_1$, the red(green) line ending with $M_{a_1a_2a_3}$ ought to be combined with the blue line to form a part of the boundary of the red(green)-blue face.
The same applies to $K_2$.
Connecting $K_1$ with $K_2$ implies therefore pasting these two red(green)-blue faces.
Summing over all possibilities of pasting two red(green) faces, we obtain the join operation.
All faces in any operator are orientable, and the pasting procedure must keep the orientation intact.
 \begin{figure}[H]
  \centering
  \includegraphics[height=3cm]{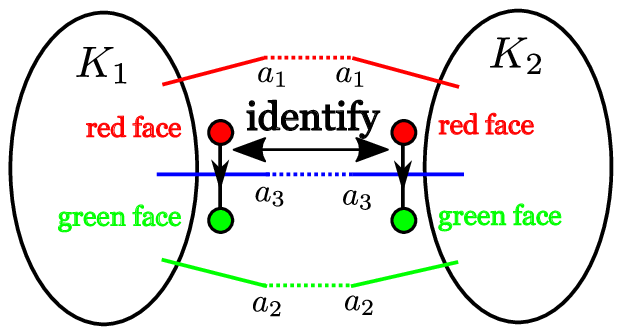}
  \caption{}\label{fig:join}
 \end{figure}%
Let us translate this into \textit{dessins}.
 In \textit{dessin},  a red(green)-blue face corresponds to its respective colored vertex, and the blue lines correspond to the red and blue vertices.
 Join is a pasting operation of two faces, and therefore, in \textit{dessin}, becomes an operation of identifying
the two sets consisting of two (respectively, red and green) vertices and the edge connecting them and summing over all possible combinations.
For example,
 \begin{align}
  &\{ K_{\cre 2}, K_{\cre 2} \} = \{ \begin{tikzpicture}
       \draw[thick] (0,0) -- (1,0);
       \filldraw [thick,fill=green] (0,0) circle (0.1);
       \filldraw [thick,fill=red] (0.5,0) circle (0.1);
       \filldraw [thick,fill=green] (1,0) circle (0.1);
       \end{tikzpicture} ~, ~
       \begin{tikzpicture}
            \draw[thick] (0,0) -- (1,0);
            \filldraw [thick,fill=green] (0,0) circle (0.1);
            \filldraw [thick,fill=red] (0.5,0) circle (0.1);
            \filldraw [thick,fill=green] (1,0) circle (0.1);
            \end{tikzpicture}  \}
    = 4 \lower3ex\hbox{\begin{tikzpicture}
         \draw[thick] (0,0) -- (0.5,0.3);
         \draw[thick] (0.5,0.3) -- (1,0);
         \draw[thick] (0.5,0.3) -- (0.5,0.9);
         \filldraw [thick,fill=green] (0,0) circle (0.1);
         \filldraw [thick,fill=red] (0.5,0.3) circle (0.1);
         \filldraw [thick,fill=green] (1,0) circle (0.1);
         \filldraw [thick,fill = green]  (0.5,0.9) circle (0.1);
         \end{tikzpicture}   }
     = 4 K_{\cre 3} , \\
   &\{K_{\cre 2}, K_{\cg 2}\} = \{ \begin{tikzpicture}
          \draw[thick] (0,0) -- (1,0);
          \filldraw [thick,fill=green] (0,0) circle (0.1);
          \filldraw [thick,fill=red] (0.5,0) circle (0.1);
          \filldraw [thick,fill=green] (1,0) circle (0.1);
          \end{tikzpicture} ~, ~
          \begin{tikzpicture}
               \draw[thick] (0,0) -- (1,0);
               \filldraw [thick,fill=red] (0,0) circle (0.1);
               \filldraw [thick,fill=green] (0.5,0) circle (0.1);
               \filldraw [thick,fill=red] (1,0) circle (0.1);
               \end{tikzpicture} \}
        = 4 ~ \begin{tikzpicture}
             \draw[thick] (0,0) -- (1.5,0);
             \filldraw [thick,fill=red] (0,0) circle (0.1);
             \filldraw [thick,fill=green] (0.5,0) circle (0.1);
             \filldraw [thick,fill=red] (1,0) circle (0.1);
            \filldraw [thick,fill=green] (1.5,0) circle (0.1);
             \end{tikzpicture}
         = 4 K_{{\cre 2}{\cg 2}}, \\
    &\{K_{\cre 2}, K_{\cb 2}\} = \{ \begin{tikzpicture}
              \draw[thick] (0,0) -- (1,0);
              \filldraw [thick,fill=green] (0,0) circle (0.1);
              \filldraw [thick,fill=red] (0.5,0) circle (0.1);
              \filldraw [thick,fill=green] (1,0) circle (0.1);
              \end{tikzpicture} ~, ~
              \lower1.2ex\hbox{\begin{tikzpicture}
                   \draw [thick] (0,0) to [out=90, in=90] (0.7,0);
                   \draw [thick] (0,0) to [out=-90, in=-90] (0.7,0);
                   \filldraw [thick,fill=red] (0,0) circle (0.1);
                   \filldraw [thick,fill=green] (0.7,0) circle (0.1);
                   \end{tikzpicture}}  \}
            = 4 ~ \lower1.2ex\hbox{\begin{tikzpicture}
                 \draw [thick] (0.5,0) to [out=90, in=90] (1.2,0);
                 \draw [thick] (0.5,0) to [out=-90, in=-90] (1.2,0);
                 \draw[thick] (0,0)--(0.5,0);
                 \filldraw [thick,fill=green] (1.2,0) circle (0.1);
                 \filldraw [thick,fill=red] (0.5,0) circle (0.1);
                 \filldraw [thick,fill=green] (0,0) circle (0.1);
                 \end{tikzpicture}}
            = 4 K_{{\cre2}{\cb 2}}.
  \end{align}
The join operation, which pastes two \textit{dessins} into a tree, is unable to produce any loop unless the original
   operator $K_{\cb 2}$ contains a loop structure.
This explains the fact that there exists operators which cannot be generated by the join operation.

On the other hand, the cut operation is defined as follows \cite{IMM1710}:
\begin{equation}\label{cut}
\Delta K = \sum_{a_1, a_2, a_3} \frac{\p^2 K }{\p M_{a_1a_2a_3} \p \bar{M}^{a_1a_2a_3}}
\end{equation}
In contrast to the join operation,  the cut operation acts on a single operator $K$ and pastes two faces which constitutes a part of the operator.
In the language of \textit{dessins},  this implies an identification of two edges of his own.
This operation of identification gives rise to a loop structure and enables us to generate operators which
 cannot be obtained from the join operation.

\begin{figure}[H]
  \centering
  \includegraphics[height=3cm]{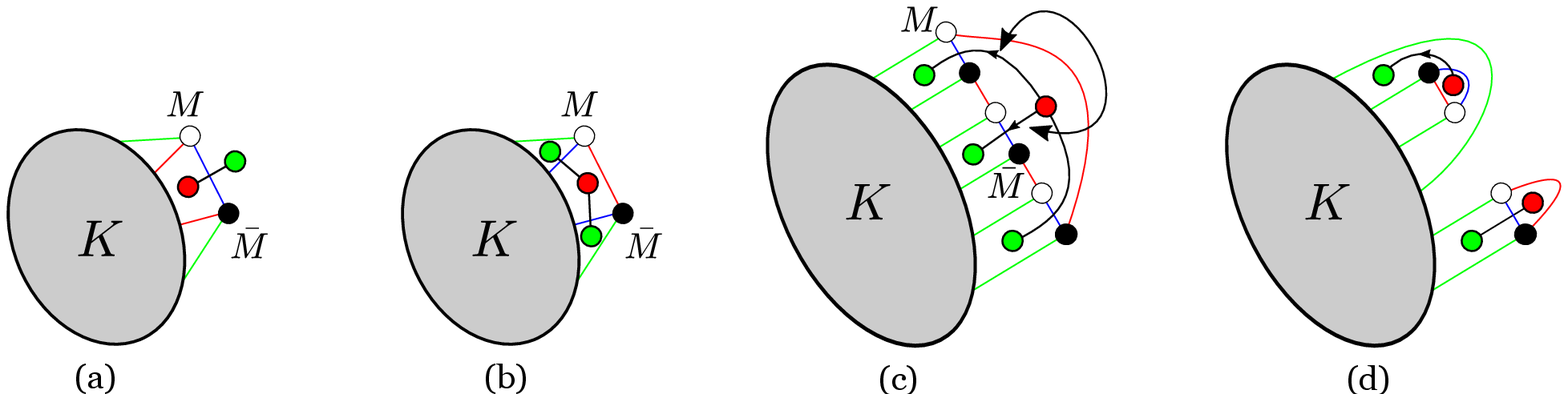}
  \caption{}\label{fig:cut}
 \end{figure}

Similarly to the join operation, the cut operation is also characterized by the identification of edges
but a consideration of several distinct cases is needed, which we further look at in some detail.
In the case when $M$ and $\bar{M}$ get differentiated  by the cut operation \eqref{cut} and
 are contracted by the pair of blue indices inside $K$, it forms the structure illustrated by
Figure \ref{fig:cut}(a).
These operators $M$ and $\bar{M}$ get eliminated, the remaining lines get connected by their color
while the blue line forms a loop giving rise to a factor of ${\cb N_b}$.
The blue line gets eliminated, which is  equivalent to the deletion of an edge in \textit{dessin}.
In the special case of a red(green) vertex isolated upon deletion of the edge,
that vertex gets eliminated as well giving rise to a factor of ${\cre N_r}$(${\cg N_g}$).

Figure \ref{fig:cut}(b) illustrates the case when the red(green) indices of
 the operators $M$ and $\bar{M}$ are contracted.
The red(green) lines give rise to a factor of ${\cre N_r}$(${\cg N_g}$), and the remaining
 green(red) and blue lines get connected.
In the language of \textit{dessins},  this case corresponds to an identification of two edges in question,
 which necessarily share a red(green) vertex, and no other edge is involved upon identification.
In accordance with the color of the vertex shared by the two edges, the cut operation
 gives rise to a factor of ${\cre N_r}$(${\cg N_b}$).

 The cut operation in the language of \textit{dessins} consists of all possible ways of  deleting an edge,
  and of all possible ways of identifying two edges with orientation and summing these over.

To summarize,  the cut operation generates the factor of ${\cb N_{b}}$ upon the deletion of an edge,
  and the factor of ${\cre N_r}$(${\cg N_g}$) upon the identification of two edges sharing a red(green) vertex.
A special care is needed in the case when $M$ and $\bar{M}$, which are eliminated by the cut,
 are not connected directly.
As an example, let us consider the situation illustrated by Figure \ref{fig:cut}(c).
Eliminating these $M$ and $\bar{M}$ operators corresponds to the operation of identifying two oriented edges in the direction indicated in the figure.
In this situation, the third edge exists and gets involved in the process of identification.
Figure \ref{fig:cut}(d) illustrates  the outcome of identification of this type.
  The factor of ${\cre N_r}({\cg N_g})$ does not appear: instead, a division of \textit{dessin} occurs.

As an example, let us look at the cut operation acting on $K_{{\cre 2}{\cb 2}{\cg 2}}$.
\begin{equation}
  \includegraphics[height=9cm]{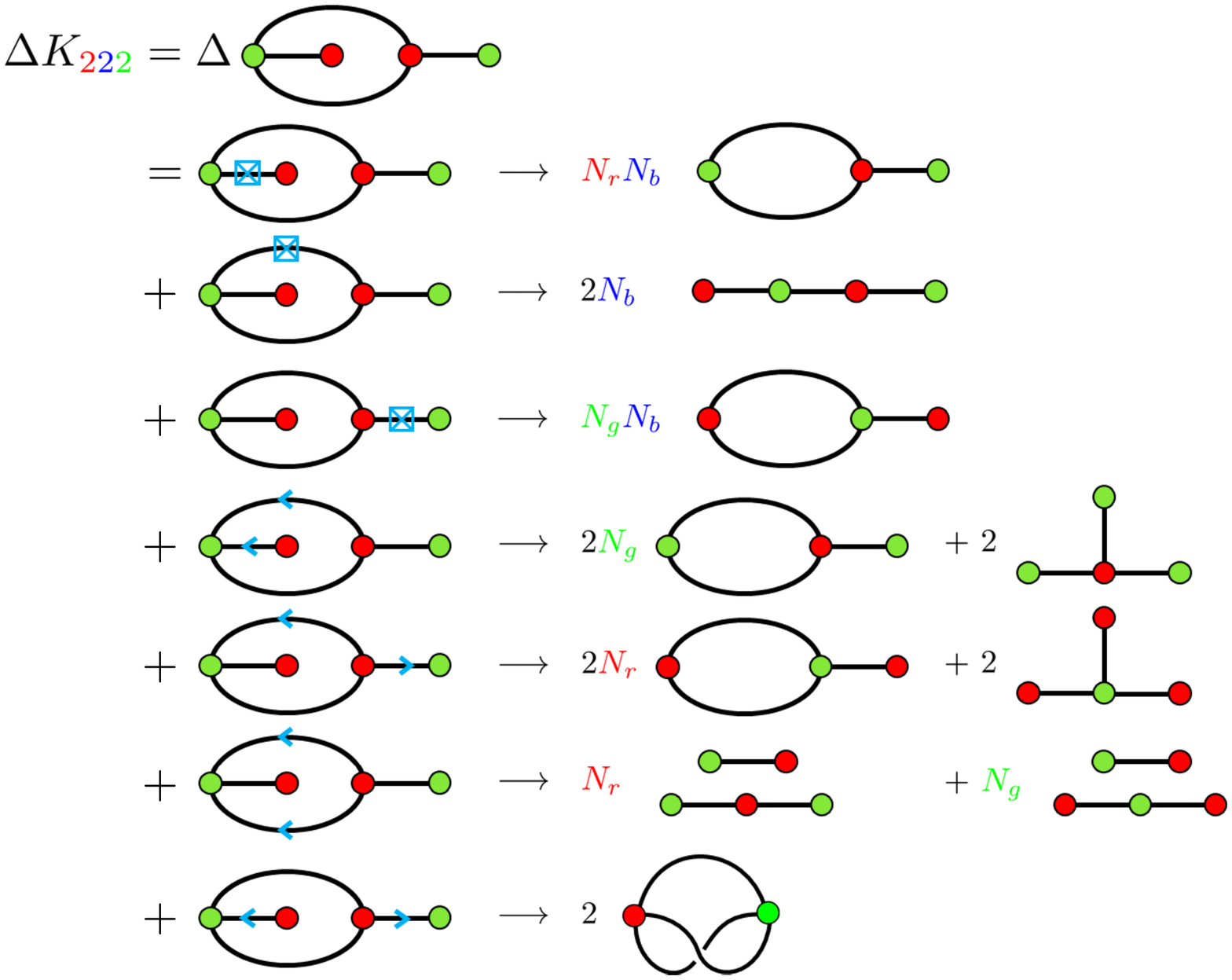}
\end{equation}
In the second equality, the first three lines are the cases of eliminating an edge
 and the remaining four lines are the cases of identifying two edges.
Here some of the terms are multiplied by factor $2$, reflecting the number of choosing one out of the two edges that constitute the loop.
On the other hand,  a nonplanar structure gets generated upon identification:
$K_{{\cre 2} {\cb 2} {\cg 2}}$ is a \textit{dessin} put onto sphere while $K_{3W}$ is a \textit{dessin} on torus.

\section*{Acknowledgments}
We thank N. Amburg and D. Vasiliev for useful discussions. 
Our work is partly supported by JSPS KAKENHI grant Number 19K03828 (H.I.) and OCAMI MEXT Joint Usage/Research Center on Mathematics and Theoretical Physics (H.I., R.Y.), by RFBR grants 19-02-00815 (A.Mor.), 19-01-00680 (A.Mir.), by
joint grants 19-51-53014-GFEN (A.Mir., A.Mor.), 21-52-52004-MNT (A.Mir., A.Mor.).

\appendix
\section{List of connected operators}\label{list}
Permutation of colors of the rank 3 operators is realized by,  in the language of the \textit{dessins},
\begin{align}
&B_{rg}([\sigma_r,\sigma_g,\sigma_b]) = [\sigma_g,\sigma_g \sigma_r \sigma_g^{-1},\sigma_b], \\
&B_{gb}([\sigma_r,\sigma_g,\sigma_b]) = [\sigma_r, \sigma_b,\sigma_b \sigma_g \sigma_b^{-1}],
\end{align}
where $B_{rg}$ exchanges the red and green colors, and $B_{gb}$ exchanges the green and blue ones.
All other color permutations can be achieved by combining $B_{rg}$ and $B_{gb}$ appropriately.
Note that $B_{rg}$ and $B_{gb}$ satisfies the braid relation,
\begin{align}
 B_{rg} B_{gb} B_{rg} = B_{gb} B_{rg} B_{gb}.
\end{align}
Exploiting it, we have composed a list of operators for levels 1-5.
In the tables, for example, an operator in the first column (RGB) transforms into  one in the second column (GRB) by exchanging the red and green colors.
The numbers in the last column describe the number of distinct operators in each row.
\begin{table}[H]
 \centering
 \caption{level 1}
  \begin{tabular}{ccccccc}
    RGB & GRB & BGR & GBR & BRG & RBG & \# \\\hline
    $K_1$ & $K_1$ & $K_1$ & $K_1$ & $K_1$ & $K_1$ &1 \\
    &&&&&&1
  \end{tabular}
\end{table}
\begin{table}[H]
 \centering
 \caption{level 2}
  \begin{tabular}{ccccccc}
    RGB & GRB & BGR & GBR & BRG & RBG & \# \\\hline
    $K_{\cre 2}$ & $K_{\cg 2}$ & $K_{\cb 2}$ & $K_{\cg 2}$ & $K_{\cb 2}$ & $K_{\cre 2}$ & 3 \\
    &&&&&& 3
  \end{tabular}
\end{table}
\begin{table}[H]
 \centering
 \caption{level 3}
  \begin{tabular}{ccccccc}
    RGB & GRB & BGR & GBR & BRG & RBG & \#\\\hline
    $K_{\cre 3}$ & $K_{\cg 3}$ & $K_{\cb 3}$ & $K_{\cg 3}$ & $K_{\cb 3}$ & $K_{\cre 3}$ &3 \\
    $K_{{\cre 2}{\cg 2}}$ & $K_{{\cre 2}{\cg 2}}$ & $K_{{\cg 2}{\cb 2}}$ & $K_{{\cg 2}{\cb 2}}$ & $K_{{\cb 2}{\cre 2}}$ & $K_{{\cb 2}{\cre 2}}$ & 3\\
    $K_{3W}$ & $K_{3W}$ & $K_{3W}$ & $K_{3W}$ & $K_{3W}$ & $K_{3W}$  &1 \\
    &&&&&& 7
  \end{tabular}
\end{table}
\begin{table}[H]
 \centering
 \caption{level 4}
  \begin{tabular}{ccccccc}
    RGB & GRB & BGR & GBR & BRG & RBG & \#\\\hline
    $K_{\cre 4}$ & $K_{\cg 4}$ & $K_{\cb 4}$ & $K_{\cg 4}$ & $K_{\cb 4}$ & $K_{\cre 4}$ &3 \\
    $K_{{\cre 3}{\cg 2}}$ & $K_{{\cg 3}{\cre 2}}$ & $K_{{\cb 3}{\cg 2}}$ & $K_{{\cg 3}{\cb 2}}$ & $K_{{\cb 3}{\cre 2}}$ & $K_{{\cre 3}{\cb 2}}$ & 6\\
    $K_{222}$ & $K_{222}$ & $K_{222}$ & $K_{222}$ & $K_{222}$ & $K_{222}$  &1 \\
    $K_{{\cre 2}{\cg 2}{\cre 2}}$ & $K_{{\cg 2}{\cre 2}{\cg 2}}$ & $K_{{\cb 2}{\cg 2}{\cb 2}}$ & $K_{{\cg 2}{\cb 2}{\cg 2}}$ & $K_{{\cb 2}{\cre 2}{\cb 2}}$ & $K_{{\cre 2}{\cb 2}{\cre 2}}$ &6 \\
    $K_{{\cre 2}{\cb 2}{\cg 2}}$ & $K_{{\cre 2}{\cb 2}{\cg 2}}$ & $K_{{\cg 2}{\cre 2}{\cb 2}}$ & $K_{{\cg 2}{\cre 2}{\cb 2}}$ & $K_{{\cb 2}{\cg 2}{\cre 2}}$ & $K_{{\cb 2}{\cg 2}{\cre 2}}$ &3 \\
    $K_{\cre 22W}$ & $K_{\cg 22W}$ & $K_{\cb 22W}$ & $K_{\cg 22W}$ & $K_{\cb 22W}$ & $K_{\cre 22W}$ &3 \\
    $K_{\cre 31W}$ & $K_{\cg 31W}$ & $K_{\cb 31W}$ & $K_{\cg 31W}$ & $K_{\cb 31W}$ & $K_{\cre 31W}$ &3 \\
    $K_{4C}$ & $K_{4C}$ & $K_{4C}$ & $K_{4C}$ & $K_{4C}$ & $K_{4C}$  &1 \\
    &&&&&& 26
  \end{tabular}
\end{table}
\begin{table}[H]
 \centering
 \caption{level 5}
  \begin{tabular}{ccccccc}
  RGB & GRB & BGR & GBR & BRG & RBG & \#\\\hline
  I & I$^{(1)}$ = XXI & I & I$^{(2)}$  & I$^{(1)}$  = XXI & I$^{(2)}$  &3 \\
  II & II$^{(1)}=$ XV & II$^{(2)}$ & II$^{(3)}$ & II$^{(4)}$ & II$^{(5)}$ &6 \\
  III & III$^{(1)}=$ XI & III$^{(2)}$ & III$^{(3)}$ & III$^{(4))}$ & III$^{(5)}$ &6 \\
  IV & IV$^{(1)}=$ IX & IV$^{(2)}$ & IV$^{(3)}$ & IV$^{(4)}$ & IV$^{(5)}$ &6\\
  V & V & V$^{(1)}$ & V$^{(1)}$ & V$^{(2)}$ & V$^{(2)}$ &3\\
  VI & VI$^{(1)}=$ XXIV & VI & VI$^{(2)}$ & VI$^{(1)}=$ XXIV & VI$^{(2)}$ &3\\
  VII & VII & VII$^{(1)}$ & VII$^{(1)}$ & VII$^{(2)}$ & VII$^{(2)}$ &3\\
  VIII & VIII$^{(1)}=$ XXII & VIII & VIII$^{(2)}$ & VIII$^{(1)}=$ XXII & VIII$^{(2)}$ &3\\
  X & X$^{(1)}=$ XXIII & X & X$^{(2)}$ & X$^{(1)}=$ XXIII & X$^{(2)}$ &3\\
  XII & XII$^{(1)}=$ XVI & XII$^{(2)}$ & XII$^{(3)}$ & XII$^{(4)}$ & XII$^{(5)}$ &6\\
  XIII & XIII$^{(1)}=$ XVII & XIII$^{(2)}$ & XIII$^{(3))}$ & XIII$^{(4)}$ & XIII$^{(5)}$ &6 \\
  XIV & XIV & XIV$^{(1)}$ & XIV$^{(1)}$ & XIV$^{(2)}$ & XIV$^{(2)}$ &3 \\
  XVIII & XVIII$^{(1)} = $ XX  & XVIII$^{(2)}$ & XVIII$^{(3)}$ & XVIII$^{(4)}$ & XVIII$^{(5)}$ &6 \\
  XIX & XIX & XIX$^{(1)}$ & XIX$^{(1)}$ & XIX$^{(2)}$ & XIX$^{(2)}$ & 3 \\
  XXV & XXV & XXV & XXV & XXV & XXV & 1 \\
  XXVI & XXVI$^{(1)}=$ XXVII & XXVI$^{(2)}=$ XXVIII & XXVI$^{(1)}=$ XXVII & XXVI$^{(2)}=$ XXVIII & XXVI &3 \\
  XXIX & XXIX$^{(1)}$ & XXIX & XXIX$^{(2)}$ & XXIX$^{(1)}$ & XXIX$^{(2)}$ &3\\
  XXIX' & XXIX'$^{(1)}$ & XXIX' & XXIX'$^{(2)}$ & XXIX'$^{(1)}$ & XXIX'$^{(2)}$ &3\\
  XXX & XXX$^{(1)}$ & XXX$^{(2)}$ & XXX$^{(3)}$ & XXX$^{(4)}$ & XXX$^{(5)}$ &6  \\
  XXX' & XXX'$^{(1)}$ & XXX'$^{(2)}$ & XXX'$^{(1)}$ & XXX'$^{(2)}$ & XXX' &3  \\
  XXXI & XXXI$^{(1)}$ & XXXI$^{(2)} = $ XXXV & XXXI$^{(3)}$ & XXXI$^{(4)}$ & XXXI$^{(5)}$ &6 \\
  XXXII & XXXII$^{(1)}$ & XXXII & XXXII$^{(2)}$ & XXXII$^{(1)}$ & XXXII$^{(2)}$ & 3 \\
  XXXIII & XXXIII$^{(1)}$ & XXXIII$^{(2)}$ & XXXIII$^{(3)}$ & XXXIII$^{(4)}$ & XXXIII$^{(5)}$ &6 \\
  XXXIV & XXXIV$^{(1)}$ & XXXIV & XXXIV$^{(2)}$ & XXXIV$^{(1)}$ & XXXIV$^{(2)}$ & 3 \\
  &&&&&&97
  \end{tabular}
\end{table}




\begin{thebibliography}{99}

\bibitem{David1985}
F.~David, ``{Planar Diagrams, Two-Dimensional Lattice Gravity and Surface
  Models},'' Nucl. Phys. B {\bf 257}, 45 (1985).

\bibitem{ADJ1991}
J.~Ambjorn, B.~Durhuus, and T.~Jonsson, ``{Three-dimensional simplicial quantum
  gravity and generalized matrix models},'' Mod. Phys. Lett. {\bf A6},
  1133--1146 (1991).

\bibitem{Sasakura1991}
N.~Sasakura, ``{Tensor model for gravity and orientability of manifold},'' Mod.
  Phys. Lett. A {\bf 6}, 2613--2624 (1991).

\bibitem{Gross1992}
M.~Gross, ``Tensor models and simplicial quantum gravity in $>$ 2-d,'' Nuc.
  Phys. B Proc. Suppl. {\bf 25}, 144--149 (1992).

\bibitem{IMM1703}
H.~Itoyama, A.~Mironov, and A.~Morozov, ``{Rainbow tensor model with enhanced
  symmetry and extreme melonic dominance},'' Phys. Lett. B {\bf 771}, 180--188
  (2017),  {{arXiv:1703.04983[hep-th]}}.

\bibitem{IMM1704}
H.~Itoyama, A.~Mironov, and A.~Morozov, ``{Ward identities and combinatorics of
  rainbow tensor models},'' JHEP {\bf 06}, 115 (2017),
  {{arXiv:1704.08648[hep-th]}}.

\bibitem{MM1706}
A.~Mironov and A.~Morozov, ``{Correlators in tensor models from character
  calculus},'' Phys. Lett. B {\bf 774}, 210--216 (2017),
  {{arXiv:1706.03667[hep-th]}}.

\bibitem{IMM1710}
H.~Itoyama, A.~Mironov, and A.~Morozov, ``{Cut and join operator ring in tensor
  models},'' Nucl. Phys. B {\bf 932}, 52--118 (2018),
  {{arXiv:1710.10027[hep-th]}}.

\bibitem{IMM1808}
H.~Itoyama, A.~Mironov, and A.~Morozov, ``{From Kronecker to tableau
  pseudo-characters in tensor models},'' Phys. Lett. B {\bf 788}, 76--81
  (2019),  {{arXiv:1808.07783[hep-th]}}.

\bibitem{IYoshi1903}
H.~Itoyama and R.~Yoshioka, ``{Generalized cut operation associated with higher
  order variation in tensor models},'' Nucl. Phys. B {\bf 945}, 114681 (2019),
  {{arXiv:1903.10276}}.

\bibitem{IMM1909} H.~Itoyama, A.~Mironov, and A.~Morozov,
``{Tensorial generalization of characters},'' JHEP {\bf 2019}, 127
  (2019),  {{arXiv:1909.06921}}.

\bibitem{IMM1910}
H.~Itoyama, A.~Mironov, and A.~Morozov, ``{Complete solution to Gaussian tensor
  model and its integrable properties},'' Phys. Lett. B {\bf 802}, 135237
  (2020),  {{arXiv:1910.03261}}.

\bibitem{MM2020}
A.~Mironov, and A.~Morozov,
  ``{Superintegrability of Kontsevich matrix model},''
Eur. Phys. J. C {\bf 81}, 270
  (2021),  {{arXiv:2011.12917}}.

\bibitem{KPT1808}
I.~R. Klebanov, F.~Popov, and G.~Tarnopolsky, ``{TASI Lectures on Large $N$
  Tensor Models},'' PoS {\bf TASI2017}, 004 (2018),  {{arXiv:1808.09434}}.

\bibitem{BGR1307}
J.~Ben~Geloun and S.~Ramgoolam, ``{Counting Tensor Model Observables and
  Branched Covers of the 2-Sphere},'' (2013),  {{arXiv:1307.6490[hep-th]}}.

\bibitem{BG2005}
J.~Ben~Geloun, ``{On the counting tensor model observables as $U(N)$ and $O(N)$
  classical invariants},'' PoS {\bf CORFU2019}, 175 (2020),
  {{arXiv:2005.01773}}.

\bibitem{BGR2106}
J.~Ben~Geloun and S.~Ramgoolam, ``{All-orders asymptotics of tensor model
  observables from symmetries of restricted partitions},'' (6 2021),
  {{arXiv:2106.01470}}.

\bibitem{MN2010}
A.~Mednykh and R.~Nedela, ``Enumeration of unrooted hypermaps of a given
  genus,'' Discrete Mathematics {\bf 310}(3), 518--526 (2010).

\bibitem{AIMMVY1911}
N.~Amburg, H.~Itoyama, Andrei Mironov, Alexei Morozov, D.~Vasiliev, and
  R.~Yoshioka, ``{Correspondence between Feynman diagrams and operators in
  quantum field theory that emerges from tensor model},'' Eur. Phys. J. C {\bf
  80}(5), 471 (2020),  {{arXiv:1911.10574}}.

\bibitem{LZ2004}
S.~K. Lando and A.~K. Zvonkin,
\newblock {\em Graphs on surfaces and their applications},
\newblock  (Springer, 2004).

\bibitem{JRRG1012}
V.~Jejjala, S.~Ramgoolam, and D.~Rodriguez-Gomez, ``{Toric CFTs, Permutation
  Triples and Belyi Pairs},'' JHEP {\bf 03}, 065 (2011),  {{arXiv:1012.2351}}.

\bibitem{AADKK0710}
N.~M. Adrianov, N.~{\relax Ya}. Amburg, V.~A. Dremov, {\relax Yu}.~{\relax Yu}.
  Kochetkov, E.~M. Kreines, {\relax Yu}.~A. Levitskaya, V.~F. Nasretdinova, and
  G.~B. Shabat, ``{Catalog of dessins d'enfants with no more than 4 edges},''
  J. Math. Sci. {\bf 158}(1), 22--80 (2009),  {{arXiv:0710.2658}}.

\bibitem{David1990}
F.~David, ``{Loop Equations and Nonperturbative Effects in Two-dimensional
  Quantum Gravity},'' Mod. Phys. Lett. A {\bf 5}, 1019--1030 (1990).

\bibitem{MM1990}
A.~Mironov and A.~Morozov, ``{On the origin of Virasoro constraints in matrix
  models: Lagrangian approach},'' Phys. Lett. B {\bf 252}, 47--52 (1990).

\bibitem{AM1990}
J.~Ambjorn and {\relax Yu}.~Makeenko, ``{Properties of Loop Equations for the
  Hermitean Matrix Model and for Two-dimensional Quantum Gravity},'' Mod. Phys.
  Lett. A {\bf 5}, 1753--1764 (1990).

\bibitem{IM1991N}
H.~Itoyama and Y.~Matsuo, ``{Noncritical Virasoro algebra of $d < 1$ matrix
  model and quantized string field},'' Phys. Lett. B {\bf 255}, 202--208
  (1991).

\bibitem{IM1991W}
H.~Itoyama and Y.~Matsuo, ``{$w_{1+\infty}$ type constraints in matrix models
  at finite $N$},'' Phys. Lett. B {\bf 262}, 233--239 (1991).

\end{thebibliography}
\end{document}